\documentclass[11pt]{article}
\usepackage{hyperref}
\usepackage{graphicx}
\pdfoutput=1
\begin{document}
\title{Particle-laden currents interacting with complex bottom topography: a numerical investigation}
\author{Mohamad M. Nasr-Azadani and Eckart Meiburg\\
\\\vspace{6pt} Department of Mechanical Engineering, \\ University of California at Santa Barbara, Santa Barbara, CA 93106, USA}
\maketitle
%% The abstract (in this file, and that submitted as text to arXiv) should include the exact phrase
%% "fluid dynamics video" or "fluid dynamics videos"
\begin{abstract}
We conduct depth-resolved three-dimensional Direct Numerical Simulations (DNS) of
bi-disperse turbidity currents interacting with complex bottom topography in the form
of a Gaussian bump.
Several flow characteristics such as suspended particle mass, instantaneous wall shear stress, transient deposit height are shown via videos. Furthermore, we investigate the influence of the obstacle on the vortical structure and sedimentation of particles by comparing the results against the same setup and but with a flat bottom surface.
We observe that the obstacle influences the deposition of coarse particles mainly in the vicinity of the obstacle due to lateral deflection, whereas for the sedimentation of  fine particles the effects of topographical features are felt further downstream.
The results shown in this fluid dynamics video help us develop a fundamental understanding of the dynamics of turbidity currents interacting with complex seafloor topography.

\end{abstract}
% main text
\section{Introduction}
Turbidity currents represent a large-scale geophysical flow phenomenon in which
the suspension travels along the bottom surface due to a horizontal hydrostatic pressure gradient (\cite{Meiburg2010}).
These currents play an important role within the global sediment cycle, and in the formation of deep-sea hydrocarbon reservoirs.
They can travel up to $O\left(1,000\ \rm{km}\right)$ along the bottom of the ocean and transport $O(\rm{km}^3)$ of sediment into deep water regions.
\\
Due to the depositional/erosional nature of turbidity currents, the interaction of such currents with seafloor topography can produce pronounced features such as channels, levees, sediment waves and gullies.
\\
In this study, we present results from DNS calculations of a bi-disperse turbidity current interacting with a complex topography in the form of a Gaussian bump. The results are obtained with our in-house code TURBINS (\cite{Nasr-Azadani2011a}), which allows us to perform large-scale simulations of these currents.
We solve the Navier-Stokes and transport equations to describe the fluid motion and evolution of particle concentration, respectively.
\section{Results}
%*******************************************************************
% Figure: 3D Configuration
%*******************************************************************
\begin{figure}
  \centerline{\includegraphics[width=0.95\textwidth]{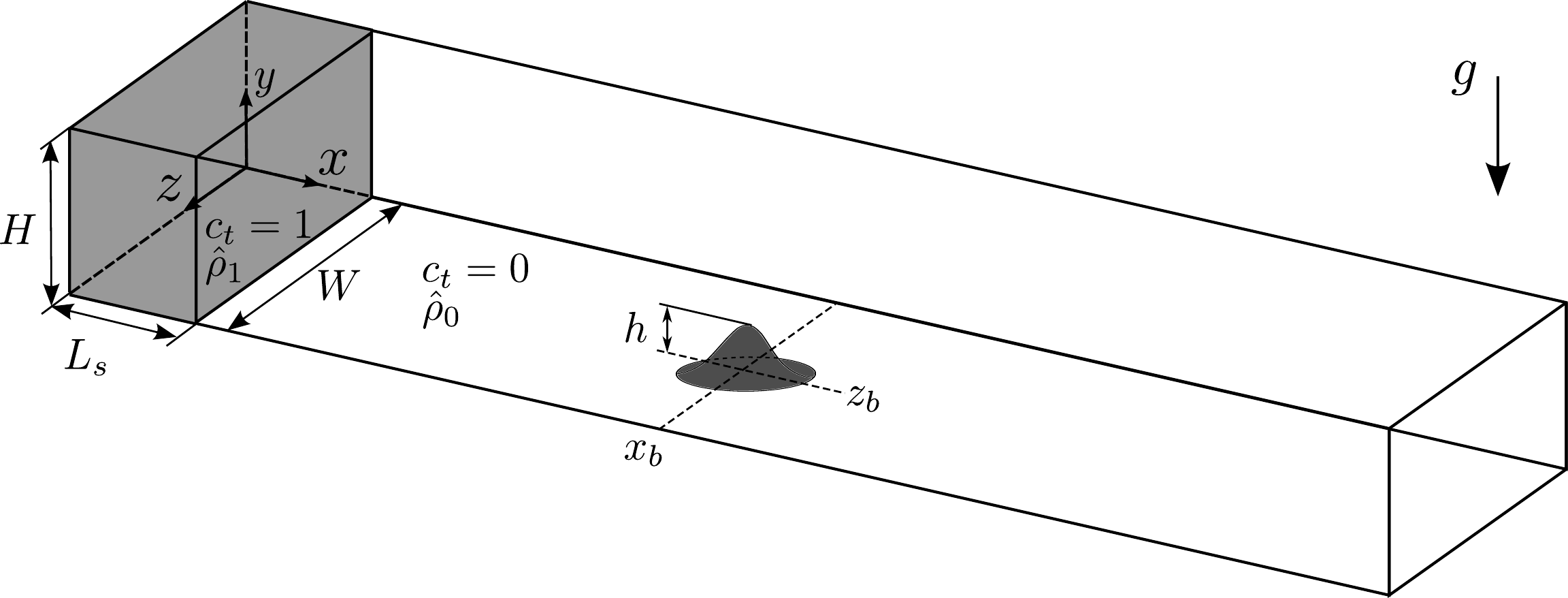}}
  \caption{Schematic of the initial set-up of the lock-exchange configuration interacting with bottom topography in the form of a Gaussian bump. The full lock has dimensions $L_s \times H \times W$ = $1 \times 2 \times 3$ and contains a bi-disperse suspension of density of $\hat{\rho}_1$. The obstacle has a height of $h=H/4$ and is centered at $(x_b,z_b) = (5.5L_s,W/2)$. At time zero, the membrane is removed and a turbidity current starts propagating along the bottom surface.}
\label{fig:configuration}
\end{figure}
%*******************************************************************
% End Figure
%******************************************************************
The suspension includes two particle sizes (see figure \ref{fig:configuration}), i.e. coarse and fine, with dimensionless settling velocities
\begin{eqnarray}
   u_s^c &=& 0.03\, , \label{eq:1} \\
   u_s^f &=& 0.006\, , \label{eq:2}
\end{eqnarray}
\noindent
and initial mass loading of 50\% and 50\%, respectively. The Reynolds number, defined by
\begin{eqnarray}
   Re = \frac{\hat{u}_b \hat{H}/2}{\hat{\nu}} \label{eq:3}
\end{eqnarray}
\noindent
is set to $Re=2,000$.
In equation (\ref{eq:3}), the buoyancy velocity $\hat{u}_b$ serves as the characteristic velocity (note: $\hat{}$ refers to any dimensional quantity).
We remark that in all results, the particle concentration $c$ refers to the volumetric particle concentration divided by the initial particle concentration in the suspension in the lock region (see figure \ref{fig:configuration}).
\\
The video presented [link here] consists of mainly two parts.
The first part presents various features of the current produced by the lock-exchange configuration interacting with the complex topography. Visualization of various flow features such as particle concentration, particle suspended mass, deposit height, and wall shear stress help us gain better insight into the flow dynamics and turbidity current characteristics. \\
We further investigate the fate of both particle sizes by tracking $48,000$ Lagrangian markers for each particle size colored by their initial location in the suspension in the lock region (see figure \ref{fig:configuration}).
From these results we observe uniform mixing and deposition of fine particles as a result of current's turbulent motion.
\\
The second section focuses on understanding the influence of the complex topography on the current.
Toward this goal, we compare our results obtained for the case with mounted obstacle against the case without any topographical feature, i.e., a flat bottom surface.
\\
As a result of the interaction of the current with the bump, we observe non-uniform vortical structures including separation regions, bi-section and re-attachment, and turbulent mixing as opposed to uniformly spaced longitudinal streaks (lobe-and-cleft instabilities) present in the flat surface simulation.
\\
Moreover, comparison of the final deposit profiles for both particle sizes suggests that coarse particles are increasingly deposited in the vicinity of the bump due to lateral deflection. For the fine particle sizes, however, we notice more pronounced differences further downstream of the Gaussian bump.

% \section*{References}
% \bibliographystyle{elsarticle-harv}
\bibliographystyle{siam}

\bibliography{References}
\end{document}